# Kinetics of segregation formation in elastic field of edge dislocation in bcc iron

Gusev A. A.[1,2], Nazarov A. V.[1,2, a]

[1] Institute for Theoretical and Experimental Physics named by A.I. Alikhanov of NRC "Kurchatov Institute", Bolshaya Cheremushkinskaya 25, 117218, Moscow, Russia
[2] National Research Nuclear University MEPhI (Moscow Engineering Physics Institute), Kashirskoe highway 31, Moscow, 115409, Russia

[a] avn46@mail.ru

We study the kinetics of the redistribution of impurity atoms in the elastic fields of dislocations by computer simulation methods. A work consists of several stages. The first is the simulation of a dislocation core structure with a Burgers vector along [100] direction by using the modified molecular static method. The second is obtaining the coefficients that determine the influence of the components of the strain tensor on the diagonal elements of the matrix of diffusion coefficients and in the equations for fluxes of carbon atoms in bcc iron. The third stage is associated with modeling the diffusion characteristics of carbon atoms by the method of molecular dynamics in a crystal without defects. The fourth and final stage uses the data about atomic structure that we obtained at first stage, as well as the characteristics calculated in the second and third, is a simulation of the formation of segregations of interstitial atoms, which is based on solving diffusion equations that take into account the elastic deformations created by the dislocation. The complex of developed models is used to analyze the kinetics of segregation formation. 3D graphs illustrate the distribution of interstitial atoms in the vicinity of a dislocation for different times at certain temperatures.

Keywords: dislocation, iron, simulation, modeling, molecular statics.

## 1. Introduction

Many properties of materials depend on the behavior of dislocations, which is largely determined by the structure of these defects, as well as by segregation of atoms in the vicinity of these defects [1-3]. In this regard, simulation of elements redistribution kinetics near linear defects is of interest for properties materials estimation.

Our approach is based on a theory of diffusion in strained metals [4-7] that was developed by us over several years, and our results of modeling of the atomic structure of a dislocation core and its surroundings [8]. We use multiscale model, which includes several stages of simulation.

The first stage that was already implemented in [8] is the study of the atomic structure of the core of the edge dislocation using the modified molecular statics method that takes into account the anisotropy of the elastic field in the medium surrounding the computational cell.
The second stage is devoted to calculating the *Strain Influence on Diffusion (SID)* coefficients [5,6] that determine the influence of the components of the strain tensor on the fluxes of carbon atoms in the bcc structure, and the diffusion characteristics of these atoms are calculated. SID coefficients are extremely sensitive to the atomic structure in the vicinity of the defect, and especially to the atomic structure in the saddle position. Therefore, for their calculation, we use the modified method of Molecular Statics (MMMS), which we developed earlier in [9-12]. The mentioned coefficients and diffusion characteristics are calculated for carbon atoms in bcc iron using the pair potentials of Johnson et al. [13,14] and EAM potential of Ackland et al. [15].

At the third stage, we model the diffusion jumps of carbon atoms in bcc iron using the molecular dynamics method and calculate the migration energy and preexponential factor for the Arrhenius equation.

The final stage is a numerical solution of the diffusion equation that determines the kinetics of the formation of impurity segregation in the vicinity of the edge dislocation. The equation takes into account the nonlinear dependence of the diffusion coefficients on strain tensor components. The values obtained in the second and third stages, as well as the strain field obtained by us earlier [8], are used to determine spatial distribution of diffusion coefficients in the vicinity of the dislocation. The redistribution of carbon atoms in the vicinity of edge dislocations with Burgers vector along [100] is simulated.

The results show that the distribution of impurity atoms is complex and the approach proposed in this work gives us an ability to study the kinetics of segregation formation in the vicinity of dislocations.

## 2. Simulation of atomic structure of edge dislocation core and it's vicinity

We use Modified Method of Molecular Static (MMMS). In this version of MS method, the main computational cell, where atoms relax, is surrounded by an elastic zone, where atoms are displaced from the ideal lattice sites in accordance with the solutions of the equations of elasticity theory. This approach, which was previously used to model point defects [9–12] and nanopores [16-18], is now adapted to find the structure of the dislocation core and its surroundings. A common feature of these models is the application of an iterative, self-consistent procedure for calculating the coordinates of atoms in the main cell and the parameters that determine the displacements of atoms immersed in an elastic medium. We believe that displacements are given in solutions of equations of the elasticity theory. Moreover, in the developed version of the model, solutions used for dislocation in the approximation of an isotropic medium, and the anisotropy is taken into account by determining the angular dependence of the Poisson coefficient and the Burgers vector determined during the iterations (now these are refined parameters).

Application of the developed model made it possible to calculate atomic displacements both in the main computational cell and in the elastic zone. A detailed description of the model and results is given in [8]. Based on the components of atomic displacement vectors obtained during simulation, the components of the strain tensor, which are used later, are calculated.

## 3. Equations for carbon flux density under strain and calculation of SID coefficients

The equations for the flux density of carbon atoms, taking into account the influence of the elastic field, were obtained by us earlier [4,6]:

$$J_1 = -\frac{1}{\Omega}\left(D_1 \frac{\partial c}{\partial x} + c \frac{\partial D_1}{\partial x}\right), \qquad J_2 = -\frac{1}{\Omega}\left(D_2 \frac{\partial c}{\partial y} + c \frac{\partial D_2}{\partial y}\right), \qquad (1)$$

where $c$ is an impurity concentration, $\Omega$ is volume per atom.
For the third axis, the equation is similar. The diagonal elements of the matrix of diffusion coefficients have the form:

$$D_1 = \frac{D}{2}\exp\left(-\frac{K_1 \varepsilon_{11}}{kT}\right)\left[\exp\left(-\frac{K_2 \varepsilon_{22} + K_3 \varepsilon_{33}}{kT}\right) + \exp\left(-\frac{K_2 \varepsilon_{33} + K_3 \varepsilon_{22}}{kT}\right)\right], \qquad (2)$$

where $D$ is the diffusion coefficient of impurity in absence of any other defects, $\varepsilon_{ij} = \frac{1}{2}(\frac{\partial u_i}{\partial x_j} + \frac{\partial u_j}{\partial x_i})$ is the strain tensor ($i, j = 1,2,3$), $u_i$ are displacement vector components, $k$ is Boltzmann constant, $T$ is absolute temperature.

$D_2$ has similar form, that one may obtain by cyclic change of indices 1,2,3.

It is easy to see that each of these coefficients depends nonlinearly on the components of the strain tensor, and this functional dependence is determined by the SID coefficients. For the case of diffusion of a carbon atom in a bcc metal (jumps of the O - T - O type), three coefficients $K_1$, $K_2$, $K_3$ determine the effect of strains on diffusion, and they are linear combinations of coefficients [4,6]:

$$K_{xx}^W = \frac{1}{2}\sum_s\sum_{k\neq s}\frac{(x_{ks}^W)^2}{R_{ks}^W}\frac{\partial E}{\partial R_{ks}}\Big|_{R_{ks}^W}, \quad K_{xx}^0 = \frac{1}{2}\sum_s\sum_{k\neq s}\frac{(x_{ks}^0)^2}{R_{ks}^0}\frac{\partial E}{\partial R_{ks}}\Big|_{R_{ks}^0}, \quad K_{xy}^W = \frac{1}{2}\sum_s\sum_{k\neq s}\frac{x_{ks}^W y_{ks}^W}{R_{ks}^W}\frac{\partial E}{\partial R_{ks}}\Big|_{R_{ks}^W} \quad (3)$$

and analogous to them. There $x_k$, $y_k$, $z_k$ are coordinates of k-th atom, $x_s$, $y_s$, $z_s$ are coordinates of s-th atom, $x_{ks} = x_k - x_s$, $y_{ks} = y_k - y_s$, $z_{ks} = z_k - z_s$, $k \neq s$, $R_{ks} = |\mathbf{r}_k - \mathbf{r}_s| = \sqrt{x_{ks}^2 + y_{ks}^2 + z_{ks}^2}$ for all atoms in system, upper index $W$ relates to system configuration at saddle position of jumping atom, upper index 0 relates to system in equilibrium state before jump occurs, $E$ is the energy of system.

For SID coefficients in the system, we have [6]:

$$K_1 = K_{xx}^W - K_{xx}^0, \quad K_2 = K_{yy}^W - K_{yy}^0, \quad K_3 = K_{yy}^W - K_{xx}^0. \quad (4)$$

At this stage, the coefficients are calculated that determine the influence of the components of the strain tensor on the flows of defects. These coefficients are very sensitive to the atomic structure of the vicinity of the defect, and especially to the positions of atoms in the saddle position. For this reason, we use the MMMS developed by us earlier [9–12], in which the atomic structure in the vicinity of the defect and the parameter that determines the displacements of atoms in the elastic zone are calculated in a self-consistent manner. A detailed description of the model originally developed for the vacancy diffusion mechanism is given in the mentioned works. In the simulation, the pair potentials of Johnson et al. [13,14] and EAM potential of Ackland et al. [15] are used. Atomic structures in the vicinity of the carbon atom located in the initial position before the jump and saddle point are found during modeling. The model made it possible to calculate not only the energy of migration of the carbon atom, but also the relaxation volumes $V_R$ and migration volumes $V_m$, and also SID coefficients $K_1$, $K_2$, $K_3$. The simulation results of this stage are given in the Table 1.

Table 1.

| Potentials | $K_1$, eV | $K_2$, eV | $K_3$, eV | $Q$, eV | $V_R / \Omega$ | $V_m/\Omega$ |
|---|---|---|---|---|---|---|
| [10,11] | 7.31 | -3.43 | 3.33 | 0.86 | 0.377 | 0.09 |
| [11,12] | 5.67 | -4.10 | 2.89 | 0.87 | 0.176 | 0.18 |

## 4. Molecular dynamics simulation of carbon atom diffusion

In carrying out this part of the work, the "natural thermostat" model [19] was used that is based on a combination of molecular dynamics (MD) methods and molecular statics, and takes into account the elastic displacements of metal atoms at large distances from the defect and the temperature dependence of these displacements. The displacements are set according to the structure of the crystal containing the defect in accordance with the spherically symmetric term of solutions from elasticity theory. At large distances from defect and taking into account thermal expansion (like in [12]):

$$u_x = C_1 \frac{x}{r^3} \quad u_y = C_1 \frac{y}{r^3}, \quad u_z = C_1 \frac{z}{r^3}, \quad C_1 = C_{10}(1 + \alpha T)^3, \tag{6}$$

where α is the coefficient of thermal expansion. The thermal expansion coefficients for various metals and the temperature dependence (6) were obtained in [21] for MD modeling, and $C_{10}$ was calculated by the MMMS in the previous section for $T = 0$.

The model gives ability to simulate various processes in crystals at constant temperature. The characteristic feature of our model reduces distortion of the fluctuation spectrum, in contrast to existing methods for modeling the canonical ensemble [20].

The calculation of the diffusion characteristics of carbon, based on the MD simulation of atomic jumps, is carried out according to the following algorithm. For each temperature, a set of time intervals that separate successive atomic jumps is obtained, the inverted intervals of these jumps are calculated, and then the average frequency Γ. This made it possible to determine the diffusion coefficient by the well-known formula [22]:

$$D = \frac{1}{6} l^2 \Gamma, \tag{7}$$

where $l$ is the length of the jump. Based on the simulation results, the temperature dependence of the diffusion coefficients is determined for two different interaction potentials of iron atoms [23]. The characteristics of diffusion coefficients and data from other works are given in Table 2.

Table 2. The migration energy of carbon atoms $E_m$ and preexponential factor $D_0$, obtained by various modeling methods and experimental data.

| Method / Potential | $D_0$, $10^{-7}$ m$^2$s$^{-1}$ | $E_m$, eV | Reference |
|---|---|---|---|
| MD / [11,12]: | 8.31 | 0.85 | This work |
| MS / [11,12] | - | 0.87 | This work |
| MD / [11,12] | 1.89 ± 0.01 | 0.71 ± 0.02 | [24] |
| MS / [11,12] | - | 0.83 | [24] |
| Experiment: | 7.0 | 0.87 | [25] |

**5. Simulation of segregation formation in vicinity of edge dislocation**

In this section, we first obtain the diffusion equation that takes into account the effect of the elastic field on the flux. To do this, we need to substitute the expressions for fluxes into the continuity equation:

$$\frac{\partial c}{\partial t} = -div \, \mathbf{J} \tag{8}$$

The redistribution of embedded carbon atoms is simulated by numerically solving the diffusion equation with allowance for the strains created by the dislocation in the elements of the matrix of diffusion coefficients (Equation 3). As a result, we find the concentration $c(x,y,t)$ in the plane normal to the dislocation line. The initial concentration of interstitial atoms is assumed to be the same and equal to $c_0$. The calculations are performed for regions of different sizes and, starting from a region of size (261 Å x 261 Å), the results are almost indistinguishable. Examples of the distribution of the concentration of carbon atoms near the dislocation are shown in Fig. 2 and 3.

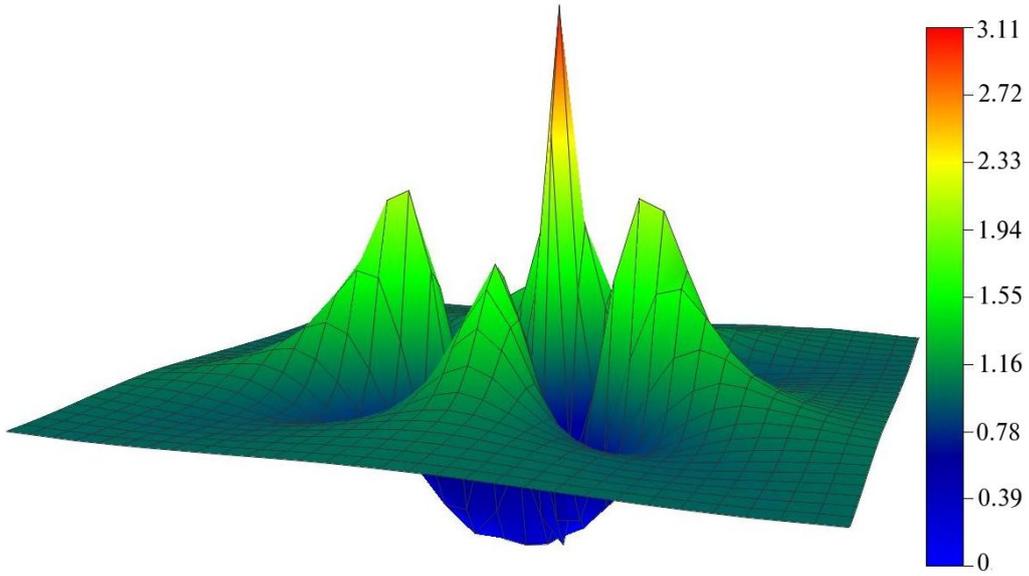

**Fig. 2.** Normalized concentration of carbon in vicinity of edge dislocation, elapsed time $t = 8$ ns, $dl = 2.86$Å, $c^{max}/c_0 = 3.11$, $T=973$K, $dl$ is the grid step, shown area is (143 Å x 143 Å).

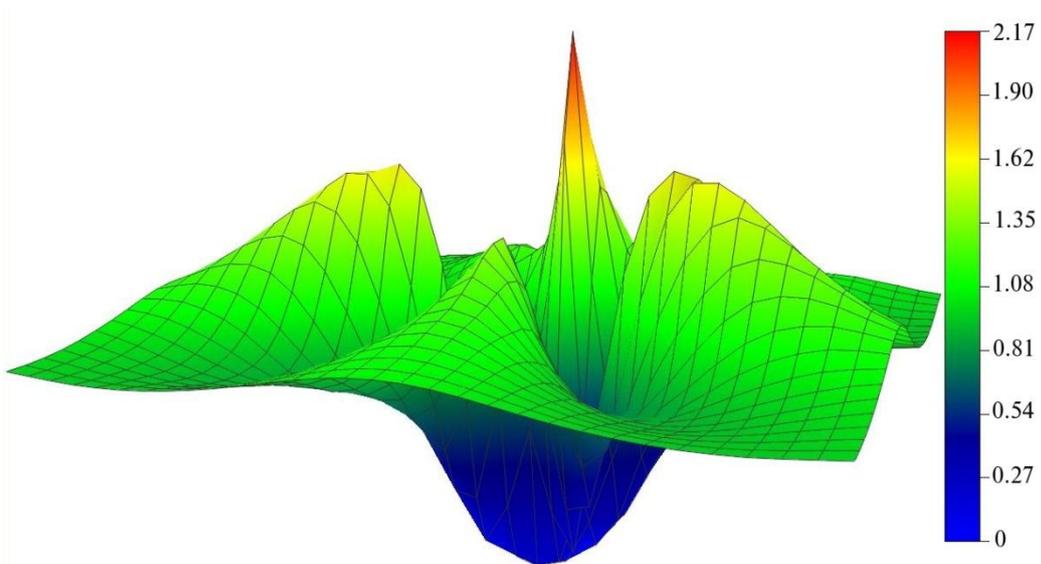

**Fig.3.** Normalized concentration of carbon in vicinity of edge dislocation, elapsed time $t = 8$ ns, $dl = 2.86$Å, $c^{max}/c_0 = 2.17$, $T=1173$K, $dl$ is the grid step, shown area is (143 Å x 143 Å).

## 6. Discussion

The simulation results show that the distribution of carbon atoms near the dislocation line has a very complex irregular character. The nature of the distribution depends on the strain field created by the dislocation, on the SID coefficients, their relationship with the activation barrier in an ideal crystal, and also on temperature. Table 3 shows spatial maximums of carbon concentrations found for different time at a temperature of $T = 1173$ K, and in Table 4 at $T = 973$ K.

Table 3. The dependence of the maximum and minimum carbon concentration on time. T = 1173K

|  | $t_1$ | $t_2$ | $t_3$ | $t_4$ | $t_5$ | $t_6$ |
|---|---|---|---|---|---|---|
| $t$, ns | 0.08 | 0.26 | 8 | 50 | 102 | 377 |
| $c^{max}/c^0$ | 2.60 | 2.80 | 2.17 | 1.62 | 1.53 | 1.42 |
| $10^6\, C_{min}/c_0$ | 145.7 | 90.8 | 42.4 | 36.5 | 31.8 | 29.8 |

Table 4. The dependence of the maximum and minimum carbon concentration on time. T = 973K

|  | $t_1$ | $t_2$ | $t_3$ | $t_4$ | $t_5$ | $t_6$ |
|---|---|---|---|---|---|---|
| $t$, ns | 0.08 | 0.26 | 8 | 50 | 102 | 377 |
| $c^{max}/c^0$ | 2.67 | 2.77 | 3.11 | 2.40 | 2.07 | 1.71 |
| $10^6\, C_{min}/c_0$ | 42.2 | 20.9 | 6.45 | 4.30 | 3.81 | 3.26 |

From the above data it follows that the concentration at the maximum increases rapidly at the beginning, and then decreases rather slowly. This feature of the formation of Cottrell atmospheres should probably be taken into account when assessing their influence on the dislocation movement.

Simulation for different initial concentrations shows that the ratio of concentrations at different coordinates to the initial one does not change. It is characteristically that the conformity in the distribution of concentrations is kept. This is why figures show relative (normalized) values.

The maximum concentration near the dislocation line noticeably exceeds the corresponding one obtained by modeling the kinetics of formation of carbon segregations in nickel [26]. In that work, for the components of the strain tensor, the results of the theory of elasticity outside the dislocation core were used, and the structure of the dislocation core was not simulated.

## 7. Conclusion

A multiscale model is developed to study the kinetics of the redistribution of impurity atoms in elastic fields of dislocations. The characteristics are calculated that determine the effect of the elastic field on the matrix of diffusion coefficients of carbon atoms in bcc iron, as well as the value of the activation barrier. Then, using the MD method, we simulated carbon migration at various temperatures and determined the migration energy and the preexponential factor for diffusion coefficients of carbon in bcc iron. In conclusion, the diffusion equation that determines the kinetics of the redistribution of carbon atoms in the vicinity of the dislocation for various temperatures is numerically solved. The simulation uses results of atomic displacements calculations in the core of dislocation and its vicinity, obtained in our work [8] and calculated therein, the components of the strain tensor.

The results indicate that the distribution of carbon concentration in the vicinity of the dislocation line has a complex nonmonotonic character and quickly reaches its maximum in the closest position to the dislocation line, then slowly decreases in this position, and the distribution minima and maxima in the vicinity of the line are irregular with complex geometry.

**Acknoledgments**

This work was supported by the Competitiveness Program of the National Research Nuclear University MEPhI (Contract No. 02.a03.21.0005).